# Self-Supervised Joint Reconstruction and Denoising of T2-Weighted PROPELLER MRI of the Lungs at 0.55T

Jingjia Chen, Ph.D.[1,2], Haoyang Pei, M.S.[1-3], Christoph Maier, M.D.[1,2], Mary Bruno, B.S.[1,2], Qiuting Wen, Ph.D.[4], Seon-Hi Shin, Ph.D.[1,2], William Moore, M.D.[1,2], Hersh Chandarana, M.D.[1,2], Li Feng, Ph.D.[1,2]

1. Bernard and Irene Schwartz Center for Biomedical Imaging, Department of Radiology, New York University Grossman School of Medicine, New York, New York, USA
2. Center for Advanced Imaging Innovation and Research (CAI$^2$R), Department of Radiology, New York University Grossman School of Medicine, New York, New York, USA
3. Department of Electrical and Computer Engineering, New York University Tandon School of Engineering, New York, USA
4. Department of Radiology and Imaging Sciences, Indiana University School of Medicine, Indianapolis, Indiana, USA

**Running title:** Self-Supervised Learning for 0.55T Lung MRI
**Word count**: 5188
**Grant support:**
NIH R01EB030549, R01EB031083, R21EB032917, and P41EB017183.
Deutsche Forschungsgemeinschaft (DFG, German Research Foundation, grant no. 512359237).

Address correspondence to:
　　Jingjia Chen, Ph.D.
　　Center for Biomedical Imaging
　　Center for Advanced Imaging Innovation and Research (CAI$^2$R)
　　New York University Grossman School of Medicine
　　227 E30th Street, Rm 758C
　　New York, NY, USA 10016
　　Telephone: +1 212-263-0182
　　Email: Jingjia.Chen@nyulangone.org

**Abstract**

**Purpose**: This study aims to improve 0.55T T2-weighted PROPELLER lung MRI through a self-supervised joint reconstruction and denoising model.

**Methods**: T2-weighted 0.55T lung MRI dataset including 44 patients with previous covid infection were used. A self-supervised learning framework was developed, where each blade of the PROPELLER acquisition was split along the readout direction into two partitions. One subset trains the unrolled reconstruction network, while the other subset is used for loss calculation, enabling self-supervised training without clean targets and leveraging matched noise statistics for denoising. For comparison, Marchenko-Pastur Principal Component Analysis (MPPCA) was performed along the coil dimension, followed by conventional parallel imaging reconstruction. The quality of the reconstructed lung MRI was assessed visually by two experienced radiologists independently.

**Results**: The proposed self-supervised model improved the clarity and structural integrity of the lung images. For cases with available CT scans, the reconstructed images demonstrated strong alignment with corresponding CT images. Additionally, the proposed model enables further scan time reduction by requiring only half the number of blades. Reader evaluations confirmed that the proposed method outperformed MPPCA-denoised images across all categories (Wilcoxon signed-rank test, $p<0.001$), with moderate inter-reader agreement (weighted Cohen's kappa=0.55; percentage of exact and within ±1 point agreement=91%).

**Conclusion**: By leveraging intrinsic structural redundancies between two disjoint splits of k-space subsets, the proposed self-supervised learning model effectively reconstructs the image while suppressing the noise for 0.55T T2-weighted lung MRI with PROPELLER sampling.

**Abbreviations**

B0: Main magnetic field.

B1+: Transmit magnetic field.

CT: Computed tomography.

UTE: Ultrashort echo time.

ZTE: Zero echo time.

FSE: Fast spin echo.

PROPELLER: Periodically rotated overlapping parallel lines with enhanced reconstruction.

SNR: Signal to noise ratio.

MPPCA: Marchenko-Pastur principal component analysis

SSL: Self-supervised learning.

DL: Deep learning.

ADAM: Adaptive moment estimator.

GRAPPA: Generalized autocalibrating partially parallel acquisitions.

FFT: Fast Fourier transform.

NUFFT: Non-uniform fast Fourier transform

CNN: Convolutional neural network.

**Introduction**

Chest computed tomography (CT) is currently the clinical standard for evaluating pulmonary anatomy and pathology thanks to its high spatial resolution, excellent air-tissue contrast, and rapid acquisition. However, a major limitation of CT is the use of ionizing radiation, which limits its applicability in young patients, pregnant women, and patients requiring repeat scans (e.g., chronic lung disease), and screening.[1,2] MRI has been considered a promising radiation-free alternative for lung imaging, but it presents many technical challenges, including low proton density in the lungs that leads to low MR signal and poor signal-to-noise ratio (SNR), the short T2* of lung parenchyma caused by numerous air-tissue interfaces, slow imaging speed, and lower achievable spatial resolution compared to CT.[3–6]

To address some of these limitations, ultrashort echo time (UTE) and zero echo time (ZTE) MRI techniques with echo times as short as a few milliseconds or nearly zero have been developed for proton-based lung imaging[7–14]. By capturing signal before significant T2* decay occurs, reducing the echo time helps preserve signal from short T2* components in the lung, thus improving the visibility of pulmonary lesions. These techniques have demonstrated great promise for morphological imaging of lung structures, particularly when combined with advanced data acquisition, image reconstruction, and motion compensation strategies. Despite these advances, UTE and ZTE MRI have not yet been adopted into routine clinical practice and remain largely confined to research settings. This is because their performance, at conventional clinical field strengths such as 1.5 T or 3 T, remains inconsistent due to various factors, including off-resonance artifacts and signal loss from B0 inhomogeneity, gradient timing errors, eddy current effects, and motion-related artifacts.[7,15]

Fast spin echo (FSE) sequences, widely available on clinical scanners, provide T2-weighted contrast and offer another promising approach for lung imaging in addition to UTE and ZTE techniques. T2-weighted FSE sequences is less sensitive to T2* decay due to the use of refocusing pulses throughout the readout. Furthermore, T2 relaxation time of lung parenchyma is much longer than T2* (e.g., T2=41 ms vs T2*=1 ms at 1.5T)[16], which allows for more robust signal retention during acquisition. Unlike UTE, which primarily yields proton density-weighted MRI contrast, T2-weighted FSE imaging provides information of tissue-specific relaxation properties. This could be particularly useful for characterizing fluid-related abnormalities such as edema, infection, or inflammation, where the presence of fluid produces high signal intensity and thus improves the characterization of pathological changes.[10,17,18] However, T2-weighted FSE imaging of the lungs remains technically demanding at conventional clinical field strengths (1.5T and 3T) and its image quality is often hindered by high specific absorption rate (SAR) from the repeated use of refocusing pulses in the echo train; transmit field (B1+) inhomogeneity that degrades the uniformity of refocusing pulses; and B0 inhomogeneity that can lead to phase errors and image artifacts.

Recently, low field MRI (< 1 T) has emerged as an exciting area of research and is now available for routine clinical use. While the reduced cost is often claimed as a major advantage, low field MRI offers viable solution for proton-based lung imaging beyond its financial benefits. In particular, the changes in relaxation times at lower field strength (longer T2 and T2*, shorter T1) are favorable for imaging the lung parenchyma[16,19–21]. The improved B1+ homogeneity at low field strength provides more uniform refocusing flip angles and the lowered resonance frequency of the proton at low field relaxes SAR constraints for T2-weighted FSE imaging[20,22]. However, the major limitation of low field MRI, particularly in lung imaging, is still the inherently low SNR. This often necessitates longer scan times and, in turn, increases its susceptibility to physiological motion that degrades the FSE signal. One established solution is to use the motion-robust T2-weighted acquisition strategy such as PROPELLER[23] (Periodically Rotated Overlapping Parallel Lines with Enhanced Reconstruction) sampling trajectory. The repetitively sampled k-space center enables motion robustness and offers signal averaging which further enhances SNR efficiency. Recent studies[22,24–26] have shown that T2-weighted PROPELLER lung MRI at 0.55T can achieve moderate to substantial agreement with chest CT in detecting and characterizing ground-glass opacities and fibrotic changes. Nevertheless, similar to other sequences, T2-weighted PROPELLER MRI at 0.55T continues to face SNR limitations, particularly in detecting subtle parenchymal abnormalities and airway structures. To address this, advanced reconstruction and post-processing techniques, such as deep learning, can be leveraged to achieve diagnostically acceptable image quality. Previous studies have explored deep learning-based PROPELLER MRI reconstruction; however, these approaches have either relied on supervised training,[27,28] or used data augmentation strategies[29] to compensate for the lack of clean, fully sampled reference images. Such methods are not directly applicable when dealing with lung images at low field, where neither fully sampled data nor noise-free images are available.

The purpose of this study was to develop a joint reconstruction and denoising technique for accelerated T2-weighted PROPELLER MRI of the lungs using self-supervised deep learning (SSL). The proposed method leverages the intrinsic redundancy of PROPELLER sampling and learns a reconstruction-denoising model directly from the undersampled noisy data without requiring fully sampled high-SNR references, which are difficult to obtain at low field strengths but are typically required in standard supervised training for deep learning (DL). We compared our approach with Marchenko-Pastur principal component analysis (MPPCA), a state-of-the-art non-DL based denoising method that has been increasingly applied in different MRI applications.[30,31] Our hypothesis was that the proposed self-supervised deep learning method could significantly improve image quality for visualizing lung structures and enable further acceleration to reduce scan times for potential clinical translation. Preliminary results were previously presented at International Society of Magnetic Resonance in Medicine (ISMRM) Annual Meeting.[32]

**Materials and Methods**

**Self-Supervised Learning for Joint MRI Reconstruction and Denoising**

Image denoising based on self-supervised learning often leverages the Noise2Noise principle[33], which assumes that two images can share the same underlying anatomical structure while containing independent noise realizations. If the noise distribution is zero-mean, one noisy image can serve as input to a neural network trained to predict the other, enabling effective denoising without requiring clean reference data. Because the noise is random, unpredictable, and uncorrelated between the two images, the network is guided to learn consistent underlying image features rather than the noise itself. This rationale enables the network to achieve a denoising effect without requiring high-SNR reference images for supervision.

In MRI, such image pairs can be generated in different ways. One straightforward approach is to acquire two repeated scans, though this doubles the acquisition time and is more practical for imaging protocols that already require multiple averages to improve SNR.[34] Another strategy is by splitting the k-space data into two disjoint subsets. Because MRI noise is independent across k-space samples, each subset can be reconstructed into an image with the same underlying anatomy but different noise realizations. However, splitting k-space essentially undersamples the acquired data, therefore the Noise2Noise training framework must be adapted to incorporate a reconstruction component. This SSL strategy has been increasingly adopted for MRI reconstruction and denoising in recent years,[35–37] especially in scenarios when fully sampled reference images are unavailable.

**Self-Supervised Learning for PROPELLER MRI**

In this study, we developed an SSL-based joint reconstruction and denoising strategy for T2-weighted PROPELLER MRI of the lungs acquired at 0.55T, where the acquisition consists of multiple blades rotated by a pre-defined angle to cover the full k-space. Let $\Lambda$ be the all-one mask covers all the sampled points in the k-space, and two compliment k-space masks $\Lambda_1$ and $\Lambda_2$ were randomly generated such that $\Lambda_1 + \Lambda_2 = \Lambda$. As illustrated in Figure 1a, these two masks were used to split the multi-coil k-space data ($y$) into two independent subsets $y_1 = \Lambda_1 \odot y$ and $y_2 = \Lambda_2 \odot y$, where $\odot$ denote elementwise Hadamard product. The first k-space subset $y_1$ is used for learning an unrolled network for reconstruction, while the second subset $y_2$ is used for loss computation to achieve self-supervision. In other words, the unrolled network is trained to produce a clean image from $y_1$ that remains consistent with the measurements in $y_2$. As $y_1$ and $y_2$ contain noise realizations of the same distribution, the reconstruction process inherently achieves a denoising effect in accordance with the Noise2Noise principle.

The reconstruction network is trained to solve the following optimization problem

$$\hat{x} = \arg\min_x \frac{1}{2}||Ex - \sqrt{W_1}y_1||_2^2 + \lambda R(x). \qquad (1)$$

Here, $E = \sqrt{W_1}F_1 C$ is the encoding operator that includes the coil sensitivity maps $C$, a non-uniform Fourier transform (NUFFT) operator $F_1$ associated with the with the first k-space mask $\Lambda_1$, and a density compensation matrix $W$. The input $y_1$ represents the first k-space split (Split-1) of the acquired multi-coil complex PROPELLER data, and $x$ is the coil-combined image to be reconstructed. The term $R(\cdot)$ denotes a regularization function with weighting parameter $\lambda$ and is modeled using a cascade of convolutional neural networks (CNNs) implemented with a U-Net architecture across unrolled iterations.

During training, a neural network $f_\theta$, parameterized by $\theta$, is optimized using a self-supervised loss defined on the second k-space split (Split-2, $y_2$) associated with mask $\Lambda_2$, which is not used in the reconstruction. The optimization objective is given by:

$$\hat{\theta} = \arg\min_\theta (E[L(F_2^H W_2 F_2 f_\theta(F_1^H W_1 y_1), F_2^H W_2 y_2)]) \qquad (2)$$

ere $F_2$ is the NUFFT operator associated with the second mask $\Lambda_2$, and $W_2$ represents its corresponding density compensation matrix. A mixed $L_1 + L_2$ loss is used for the loss function $L(.,.)$, which enforces consistency between the reconstructed image and the unseen k-space data from Split-2.

**Network Implementation**

As described above, the acquired PROPELLER k-space data is split into two disjoint partitions, Split-1 ($y_1$) is used to train the unrolled reconstruction network, while Split-2 ($y_2$) is used to compute the training loss $L$. To ensure robustness and generalization, the k-space splitting ratio is randomly selected between 0.3 and 0.99 for each image and is varied dynamically across all training epochs.

Coil sensitivity maps are jointly estimated during network training through a CNN branch, as implemented in prior works.[38] Specifically, a U-Net architecture is used to estimate coil sensitivity maps from the center of the PROPELLER k-space and is optimized jointly with the unrolled reconstruction network $f_\theta$ using the same loss function $L$ described in Equation 2.

The unrolled network consists of 12 cascaded iteration blocks and uses the ADAM[39] optimizer with a learning rate of 1e-4, a batch size of 1, and approximately 8.66 million trainable parameters. The model was trained for 200 epochs using PyTorch (v2.0) on a high-performance workstation equipped with an NVIDIA Tesla A100 GPU.

The inference process is shown in Figure 1b. Once training is complete, inference can be performed directly on the full PROPELLER k-space data without splitting to enable reconstruction of denoised, coil-combined images. In this setting, $y_1$ in Equation 1 is replaced with the full k-space $y$ without applying any masking, and the NUFFT operator $F$ incorporates the complete acquisition trajectory without splitting, which gives $\Lambda_1 = \Lambda$, and $\Lambda_2 = \emptyset$ (empty set).

**Comparison with MPPCA**

To evaluate the performance of our proposed SSL-based joint reconstruction and denoising framework, we compared our approach against MPPCA, a non-deep learning denoising technique.[30,31] MPPCA denoises MRI data by analyzing local spatiotemporal patches and applying singular value decomposition (SVD) to identify low-rank signal components. It removes noise based on the Marchenko-Pastur distribution, which statistically models the singular value spectrum of random Gaussian matrices. The MPPCA denoising method was selected as a reference for comparison because it has demonstrated great promise in denoising low-SNR MRI data and has been increasingly used in scenarios with inter-image redundancy, such as diffusion-weighted imaging[40] and functional MRI.[41] In our study, MPPCA was applied as a post-reconstruction denoising method to the PROPELLER datasets. Since our datasets consist of static images without a temporal dimension, instead, we applied MPPCA along the coil dimension, treating the multi-coil images as a pseudo-dynamic series.

**Imaging Experiments**

This HIPAA-compliant and IRB-approved study included 44 patients (23 women and 21 men, mean age = 51.54 ± 13.24 years) who were followed after COVID-19 infection. Written informed consent was obtained from all participants prior to the imaging exams. All MRI scans were performed between March 2022 and August 2023 on a prototype 0.55T MRI scanner (ramped-down Aera, Siemens Healthineers). All subjects underwent free-breathing, respiratory-triggered T2-weighted PROPELLER MRI of the lungs in the axial plane. Relevant imaging parameters included: field of view (FOV) = 380 × 380 mm$^2$, matrix size = 320 × 320, slice thickness = 5 mm, number of acquired slices = 40 slices, TE = 65 ms, TR = 2000 ms, flip angle (FA) = 170º, and 18 blades per slice for fully sampled acquisition. GRAPPA (Generalized Autocalibrating Partially Parallel Acquisitions) acceleration with a factor of 2 (R = 2) was applied along the phase-encoding direction for each blade (referred to as in-blade undersampling). All data were acquired during free-breathing with respiratory triggering. The acquisition time was 4 minutes and 45 seconds, and the total elapsed scan time depends on individual subject's respiration pattern. Corresponding chest CT images for four subjects acquired within three months of the MRI scans were available for comparison.

All MR datasets were reconstructed offline using three different methods as described below. Image reconstructions were performed directly on the acquired data with a fixed in-blade acceleration were referred to as R=2. Additionally, we also performed reconstruction with reduced number of blades, which combined 2-fold in-blade and 2-fold cross-blade undersampling, achieving a total acceleration of R=4.

1. **Standard GRAPPA Reconstruction (R = 2):** This reconstruction scheme implements the procedure used for standard reconstruction on the scanner. GRAPPA reconstruction was applied to each blade separately, followed by NUFFT to combine all the blades into a full image. To correct for inter-blade motion, phase correction was performed on each blade using the method described in previous study.[23] Coil sensitivity maps were estimated using the Walsh method[42] and were used to combine the multi-coil data into a final image.

2. **MPPCA + GRAPPA Reconstruction (R = 2 and R=4):** In this reconstruction method, MPPCA denoising was applied prior to GRAPPA reconstruction. Specifically, denoising was performed on reduced FOV images generated by zeroing out the unacquired k-space samples. As shown in Figure 2, a coil covariance matrix was first estimated from a noise pre-scan and used to decorrelate noise across coil elements. A fast Fourier transform (FFT) was then applied blade by blade to convert the acquired k-space data into the image domain. Since each blade contains in-blade 2X acceleration, the image of each blade contains aliasing due to skipped k-space lines. Then, for both GRAPPA reference scan and imaging data of each blade, MPPCA denoising was performed along the coil dimension. After denoising, the multicoil images of each blade were transformed back into k-space using inverse FFT. Standard

GRAPPA reconstruction was then applied for the denoised k-space data as described in the first method, followed by NUFFT and coil combination to generate the final image.

3. **Self-supervised Joint Denoising and Reconstruction (SSL, R = 2 and R = 4)**: This method follows the SSL framework as described in Figure 1. A total of 29 datasets were used for training and validation, while the remaining 15 were used for evaluation. The model was trained on fully acquired 18-blade datasets with 2-fold acceleration and learned to jointly denoise and reconstruct undersampled data directly from the PROPELLER k-space. To simulate higher acceleration, a retrospective 4-fold acceleration was achieved by uniformly subsampling every other blade from the original 18-blade acquisition (cross-blade undersampling), resulting in 9 blades per slice. The same SSL pipeline was applied, using 29 datasets for training and validation and the remaining 15 for evaluation.

**Image Quality Assessments**

Two experienced radiologists with 21 and 6 years of chest imaging reading experience independently assessed images quality using a 5-point Likert scale. All images reconstructed using the four methods described above were pooled and randomized for blinded evaluation. Images reconstructed using standard GRAPPA reconstruction, with MPPCA denoising, and with the proposed SSL approach (including R=2 and R=4 settings) were scored. Although we also reconstructed the image with R=4 acceleration with MPPCA, the images were already noticeably worse in terms of the quality and noise level and therefore were not sent for scoring. Each image was scored for six assessment categories, including overall image quality, perceived noise level, visualization of great vessels (i.e., aorta and pulmonary artery), large airways, segmental arteries, and segmental bronchi. The scoring criteria were as follows: 1=not readable, non-diagnostic quality or structure not visible; 2=high noise level with limited but acceptable visualization; 3=moderate noise with marginally restricted assessment; 4=low noise with good visualization and unrestricted assessment; 5= excellent quality with no noticeable noise.

**Statistical Analysis**

Statistical analysis was performed using R (version R 4.5.1 for macOS) with "rstatix" and "Metrics" libraries. Freidman test was performed as the overall test, followed by one-tail Wilcoxon signed-rank test for the following pairs: SSL (R=2) vs GRAPPA; SSL (R=4) vs GRAPPA; SSL (R=2) vs MPPCA; SSL (R=4) vs MPPCA. P-values were adjusted for multiple comparisons using the Benjamini-Yekutieli method. Inter-reader agreement was evaluated using quadratic weighted

Cohen's kappa, which was interpreted as: < 0.00 = poor, 0.00-0.20 = slight, 0.21-0.40 = fair, 0.41-0.60 = moderate, 0.61-0.80 = substantial, and 0.81-1.00 = almost perfect agreement;[43] The percentage of exact agreement and agreement within ±1 point was also calculated[44].

**Results**

Figure 3 shows a representative slice from a subject without morphologic abnormality. While MPPCA effectively reduces overall noise, the proposed SSL-based method further suppresses noise and improves visualization of lung parenchyma structures. In particular, the SSL model trained at R = 4, using only 9 blades uniformly selected from the original 18 blades, achieved a twofold reduction in acquisition time without compromising image quality. The difference map (scaled up by 5-fold) confirms that the SSL-based method effectively removes noise while preserving anatomical detail. In addition, the SSL method effectively addresses spatially varying noise, as highlighted by the yellow circle. Aliasing artifacts, indicated by the green arrow, are also eliminated by the SSL method but remain visible in the GRAPPA and MPPCA reconstructions.

Figure 4 presents two axial slices from another subject with no visible pathology. While MPPCA enhances perceived SNR, noise suppression is mainly limited to the background with little effect in the lung regions in this case. In contrast, our SSL-based method yields visibly sharper anatomical detail and lower noise, even at R=4 acceleration rate. Aliasing artifacts in the background are also reduced with SSL-based method.

Results from Figures 3 and 4 suggest that the performance of MPPCA is less consistent compared to the SSL-based method. Figure 5 further highlights this difference across different cases using R=2 dataset (with all 18 blades). For subject 1, the MPPCA reduces noise level and recovers fine vasculature inside the lung, as pointed by the green arrow. However, in subject 2, the MPPCA result (especially at the anatomy pointed by the red arrow) visually is very similar to the GRAPPA reconstruction as if the MPPCA was not being applied. In both cases, the SSL-based method provides superior image quality compared to GRAPPA and GRAPPA+MPPCA. For Subject 3 in Figure 5, residual aliasing artifacts from undersampling, pointed by the yellow arrow, intrude into the lung parenchyma and become more pronounced after MPPCA, indicating that denoising alone is insufficient to correct reconstruction errors inherent to GRAPPA.

Figures 6-8 present three cases with pulmonary pathology in comparison with corresponding CT images. In Figure 6, a right upper lobe cyst (pneumatocele) confirmed by CT (red arrow) is barely visible on the GRAPPA image. The lesion becomes more apparent with MPPCA and is clearly delineated in the SSL images at both R = 2 and R = 4. Minor differences between the SSL images at R = 2 and R = 4 likely reflect the variation of respiratory motion states in these two images due to different scan durations at different acceleration rates. Figure 7 shows a case with subpleural lines and subtle ground-glass opacities (red arrow), which are blurred on both GRAPPA and MPPCA images but better visualized with SSL, which provides the sharpest delineation at both acceleration rates. Figure 8 presents a case with a mosaic attenuation pattern. Regions of pathological hypodensity on CT correspond to hypointense signal areas in all MRI images, but these areas are best visualized with SSL at both R = 2 and R = 4.

Figure 9 summarizes the image quality scores along with the minimum and maximum scores averaged from two independent readers across six evaluation categories for different methods. Across all categories, SSL-based reconstruction and denoising, both at R=2 and R=4, consistently received significantly higher scores than conventional methods. The mean score ($\pm$ standard deviation) for the perceived noise level for GRAPPA reconstruction is 2.63 ($\pm 0.48$), for GRAPPA+MPPCA reconstruction is 3.00 ($\pm 0.46$), for SSL (R=2) is 4.80 ($\pm 0.32$) and for SSL (R=4) is 4.67 ($\pm 0.41$). The mean score ($\pm$ standard deviation) for the overall image quality for GRAPPA reconstruction is 2.80 ($\pm 0.49$), for GRAPPA+MPPCA reconstruction is 2.90 ($\pm 0.60$), for SSL (R=2) is 4.00 ($\pm 0.53$) and for SSL (R=4) is 3.87 ($\pm 0.58$). The Wilcoxon signed-rank test show p<0.001 significance for SSL(R=2) vs GRAPPA, SSL(R=2) vs GRAPPA+MPPCA, SSL(R=2) vs GRAPPA and SSL(R=4) vs GRAPPA+MPPCA in both perceived noise level and overall image quality scoring categories. Detailed test statistics are included in Supporting Materials Table S1, S2 and S3, Inter-reader agreement was moderate, with a quadratic-weighted Cohen's kappa of 0.55. The percentage of exact and within ±1 point agreement was 91%. Despite significant image quality improvements using SSL reconstruction, the visualization of small bronchial structures remains limited, likely reflecting an intrinsic limitation in the spatial resolution of T2-weighted lung MRI.

Summarizing from all the data, the average inference time for SSL-based reconstruction and denoising was 13.2 seconds per case for the original datasets with an in-blade acceleration factor of R=2, and 12.4 seconds per case for the retrospectively accelerated dataset using half the blades at an acceleration factor of R=4.

Finally, Figure 10 shows axial slices of the liver from two subjects, which were partially included in the PROPELLER scans. Consistent with previous observations, MPPCA denoising along the coil dimension reduces noise to a limited extent. In contrast, the SSL-based joint reconstruction and denoising approach effectively suppresses noise, removes residual undersampling artifacts, and improves the depiction of fine liver structures. Compared to the lung images, the SSL method appears to perform better in the liver, likely due to the higher proton density of liver tissue. These findings suggest that the proposed SSL approach may also be applicable to other organs beyond the lungs.

**Discussion**

Lung MRI has long been a topic of research interest, and its radiation-free nature makes it especially appealing for pediatric imaging, and repeated imaging, including lung disease screening and ongoing monitoring. However, unlike MRI for other organs, MRI of the lungs has not made its way into widespread clinical adoption due to low intrinsic SNR, large B0 inhomogeneity caused by air-tissue magnetic susceptibility differences in the lungs, and consequently prolonged scan times resulting from the need to mitigate these limitations.

In this work, we present a joint denoising and reconstruction framework for respiration-triggered T2-weighted PROPELLER lung MRI at 0.55T, leveraging a self-supervised learning strategy that explores the intrinsic redundancy within the data itself when fully sampled or clean reference is not available. During training, the undersampled k-space is randomly partitioned into two disjoint subsets, one as input for the unrolled network, and the other for loss evaluation, allowing the model to learn from within the real-world noisy data itself. The proposed method improves image quality and SNR through joint reconstruction and denoising, while enabling more aggressive undersampling on top of the acceleration already achieved by GRAPPA. Specifically, we demonstrate that the number of blades can be halved while maintaining diagnostic image quality, effectively reducing the total scan time to approximately 2-3 minutes. This substantial acceleration, combined with improved structural clarity, facilitates low-field lung MRI for clinical lung diagnostic imaging when repeated imaging is needed or for vulnerable population such as young children and pregnant female.

Improving SNR and reducing scan time are inherently conflicting goals in MRI, as shorter acquisitions typically result in increased noise and reduced image quality. However, by jointly performing denoising and reconstruction, our method addresses both challenges simultaneously. Instead of treating denoising as a post-processing step, the self-supervised framework integrates noise suppression directly into the image reconstruction process through the assumption that the two splits shared the same underlying structural information and different noise realization of the same distribution. As a results, the network learns to predict the consistent structural features while suppressing noise, which cannot be learned due to its random nature[33]. This joint formulation allows for more aggressive undersampling while preserving, and even enhancing, diagnostic image quality—offering a practical path toward faster, high-quality lung MRI at low field strength.

In terms of improving SNR, in this work, we first explored the application of MPPCA, which was originally developed for removing additive white Gaussian noise by leveraging intrinsic redundancy along the dynamic dimension of diffusion MRI data. In our case, the reconstruction task involves a static T2-weighted image, so we instead rely on redundancy across the coil dimension. In our experiments, we found that applying MPPCA along the coil dimension can be effective in some cases but does not yield consistent results across all cases. This is likely because the data redundancy along the coil dimension is not always guaranteed and depends on the degree of overlap between coil sensitivity regions. For example, in an extreme case, if the coil sensitivities are completely disjoint, there is no shared signal to exploit, and thus no redundancy for MPPCA to leverage. At the other extreme, if the coils have identical coverage and full overlap, the redundancy would be almost analogous to that in the diffusion MRI case. In practice, the level of coil overlap varies depending on the subject's anatomy and coil configuration, leading to variability in the effectiveness of MPPCA and making its performance less robust across patients.

MPPCA removes noise components based on the Marchenko-Pastur distribution of singular values of random matrices requiring Gaussian distribution of the noise and relies on sufficient separation between signal and noise singular values, which a condition that becomes difficult to satisfy at very low SNR. Our self-supervised approach does not rely on those assumptions but only requires the two data partitions to have zero-mean noise with matching statistics. The SSL framework in its early developments[35,45] were designed for image reconstruction from undersampled Cartesian MRI data. In our adaptation to non-Cartesian MRI at low field, where noise levels are substantially higher, we observed strong denoising effects alongside successful reconstruction. As analyzed before, SSL splitting two parts along the readout direction satisfies the Noise2Noise[33] framework and therefore provides denoising power. It is important to note that the split along the readout direction is not intended for acceleration, but rather to create two subsets of k-space data that share identical noise and signal statistics—one for the training branch and one for loss evaluation. In our implementation, the k-space splitting ratio is randomly varied between 0.3 and 0.99 during training, which allows the network to be exposed to the entire view of the k-space. This design enables inference using the full k-space data, without needing to discard any portion for the sake of maintaining the self-supervised learning structure.

For the scan time reduction, we tested the performance of this proposed SSL framework when dealing with even more aggressively undersampled setting by retrospectively removing number of blades of each image. The original dataset is already acquired at R=2 acceleration through 2X GRAPPA along the phase encoding direction (i.e. the width of the blade). By halving the number of blades, we achieved additional 2X acceleration, resulting in an acceleration factor R=4. In PROPELLER acquisitions, in-blade undersampling does not substantially reduce scan time, as it only shortens the echo train length within each blade. In contrast, reducing the number of blades directly reduces the number of TRs, thereby effectively decreasing the total scan time. When directly comparing R=2 and R=4 reconstruction, the noise level remains similarly low; however slight structural differences were observed. Such variations were also present in the noisy MPPCA results when comparing R=2 and R=4 setting, confirming that the differences are presumably intrinsic to the data rather than reconstruction artifact. A plausible explanation is that residual variability within the respiratory gating window leads to slight differences in respiratory state across blades, resulting in subtle variations when averaging over different subsets of blades. Reassuringly, the reader study found no significant difference in image quality scores between the R = 2 and R = 4 results, indicating that the proposed method maintains diagnostic quality even at higher acceleration.

In addition to the denoising and acceleration benefit, the SSL method learns the reconstruction process and shows more robustness than conventional non-DL GRAPPA reconstruction in terms of aliasing artifacts reduction. This is likely because GRAPPA reconstruction is known to amplify noise when autocalibration lines have low SNR ,[46] which is the case in low field lung imaging. In contrast, the SSL method does not rely on linear kernel fitting and interpolation from noisy autocalibration lines, making it inherently more resilient to k-space interpolation errors.

There are a few limitations to consider when adapting this study for further and broader applications. First, due to the retrospective design, we did not have access to multiple signal averages that could potentially serve as a reference or ground truth for evaluation. Indeed, this is precisely why SSL is particularly valuable in such scenarios. Second, our study uses a relatively small cohort size with 44 patients scanned on a prototype low-field system. Although the visual results are promising, this limited cohort size (29 datasets for training and 15 datasets for validation) prevents us from making stronger statistical conclusions regarding the broader adoption of the proposed method. Even so, our reader study demonstrates significant improvements in image quality and shows that T2-weighted lung MRI can be excellent in visualizing great vessels, large airways when reconstructing using SSL framework, providing initial evidence supporting the clinical value of our approach. Future study should include more cases with various pathologies and paired with ground truth CT as references for diagnosis performance evaluation.

Despite these limitations, our work highlights several promising directions for applications. From a technical perspective, our framework offers a practical solution for self-supervised joint denoising, and reconstruction tailored for low-field, non-Cartesian MRI and it does not rely on external reference data. Clinically, the proposed method supports broader adoption of lung MRI by offering a radiation-free alternative to CT, particularly in scenarios requiring repeated imaging or for vulnerable cohorts such as pediatric patients and pregnant female patients. Although this study focuses on the lung MRI, in theory, this proposed approach is agonistic to the organ. As observed within the field of view, at regions with higher intrinsic SNR, such as the liver, the image quality improvement using SSL is even more prominent.

In conclusion, we propose a self-supervised model for the joint denoising and reconstruction of the T2-weighted PROPELLER lung MRI at 0.55T without the need for external training labels. Compared to standard GRAPPA reconstruction, and GRAPPA reconstruction with MPPCA denoising, the proposed method effectively removes noise from low-SNR lung images and recovers the detailed anatomical structures in the lungs that were obscured by noise. We further demonstrated that this method supports an additional 2-fold acceleration for reducing scan time. Other regions, such as the liver, may also benefit from the proposed method for jointly denoising and reconstructing images at 0.55T.

**Acknowledgments**

This work was supported in part by the NIH (R01EB030549, R01EB031083, R21EB032917, and P41EB017183) as well as by the Deutsche Forschungsgemeinschaft (DFG, German Research Foundation, grant no. 512359237) and was performed under the rubric of the Center for Advanced Imaging Innovation and Research (CAI$^2$R), an NIBIB Biomedical Technology Resource Center. The authors would like to thank Mahesh Keerthivasan for their support with data acquisition.

**Figures:**

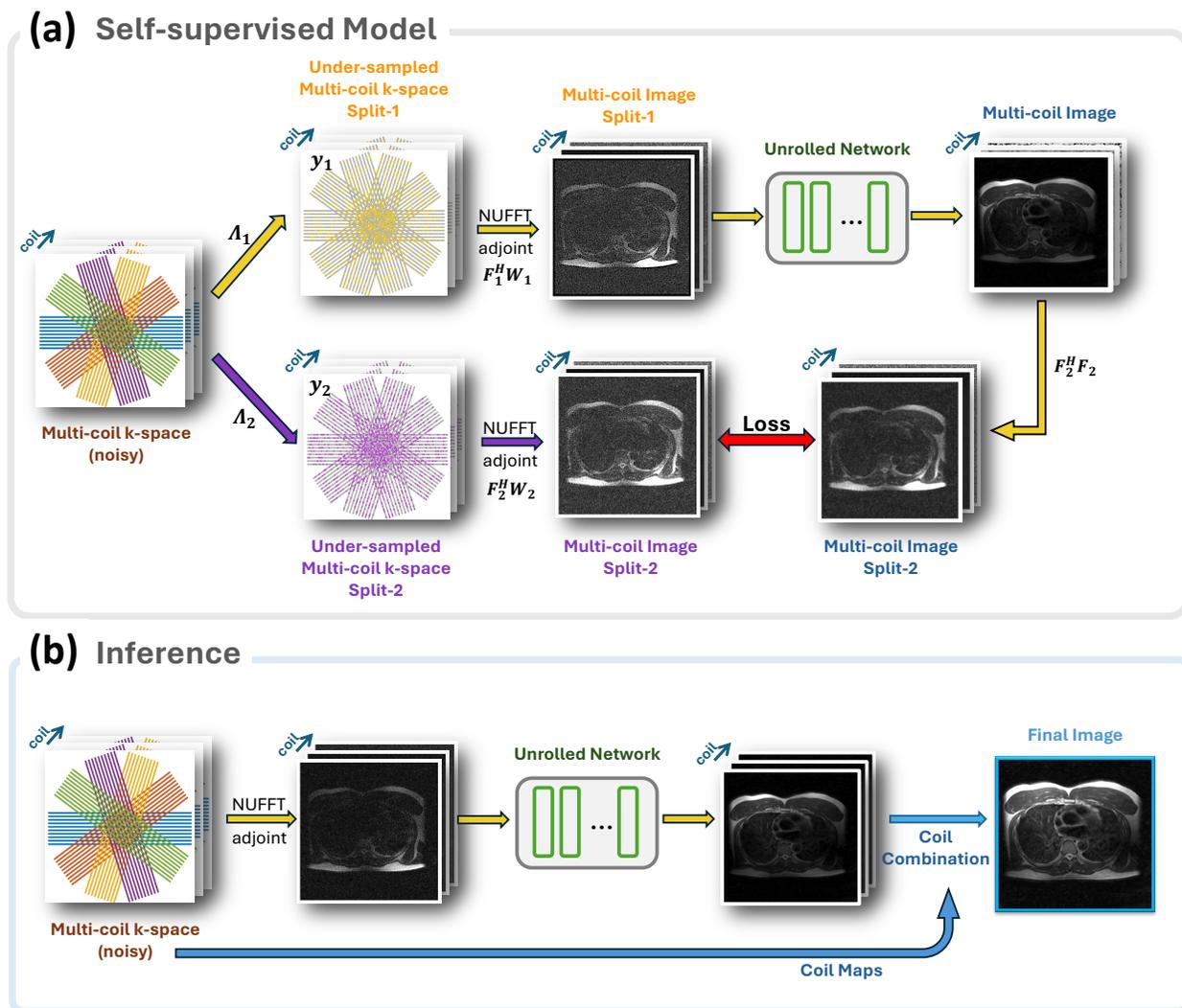

**Figure 1**. Illustration of the self-supervised joint denoising and reconstruction pipeline. (a) The training. The undersampled k-space data are split along the readout direction into two disjoint subsets $y_1$ and $y_2$ with various ratios. The first subset is used as input to an unrolled network to generate a reconstructed k-space output, which is then compared against the second subset to compute the loss. This self-supervised strategy enables the network to learn directly from the undersampled data without requiring fully sampled reference images. (b) The inference. The entire acquired k-space data, without splitting, is used for inference for reconstruction multicoil clean results. Coil sensitivity maps are calculated through the center of the original acquired k-space data.

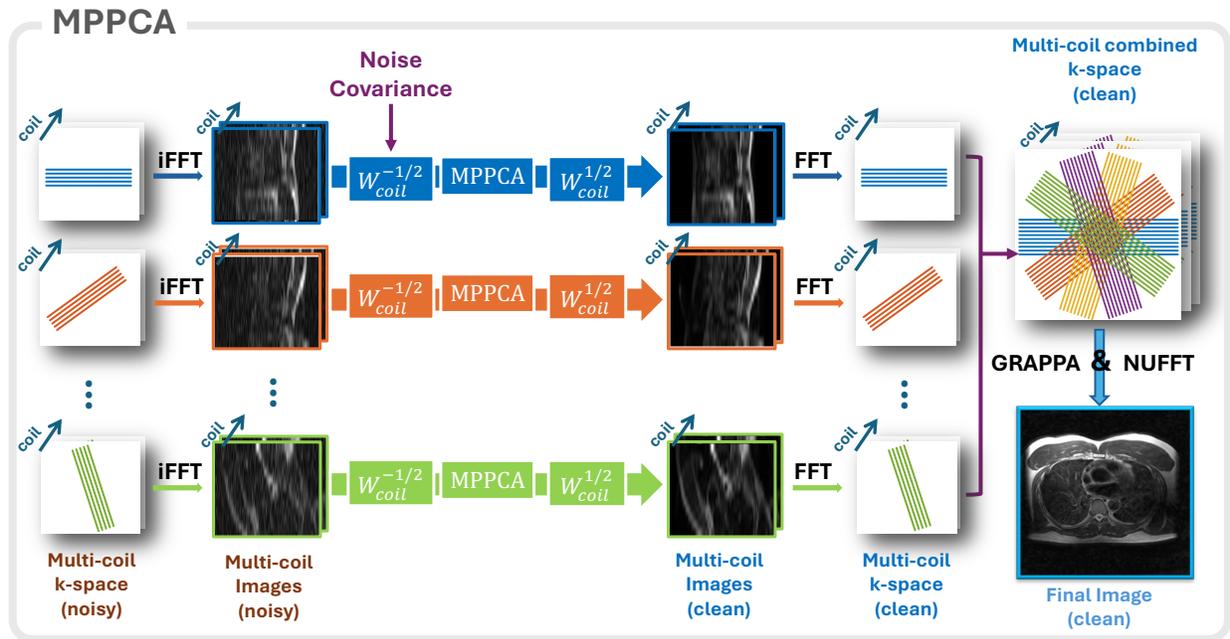

**Figure 2**. Illustration of the integration of MPPCA denoising into the GRAPPA reconstruction pipeline. K-space data from each blade are denoised separately. First, an iFFT is applied to bring the multi-coil k-space blade into the image domain. The noise covariance matrix is then used to decorrelate the multi-coil image, after which MPPCA denoising is applied along the coil dimension. The decorrelation process is subsequently reversed, and the denoised multi-coil image is transformed back into k-space. GRAPPA is then applied to fill in the missing k-space lines, followed by NUFFT to generate the final image. GRAPPA: Generalized autocalibrating partially parallel acquisitions. MPPCA: Marchenko-Pastur principal component analysis. NUFFT: Non-uniform fast Fourier transform.

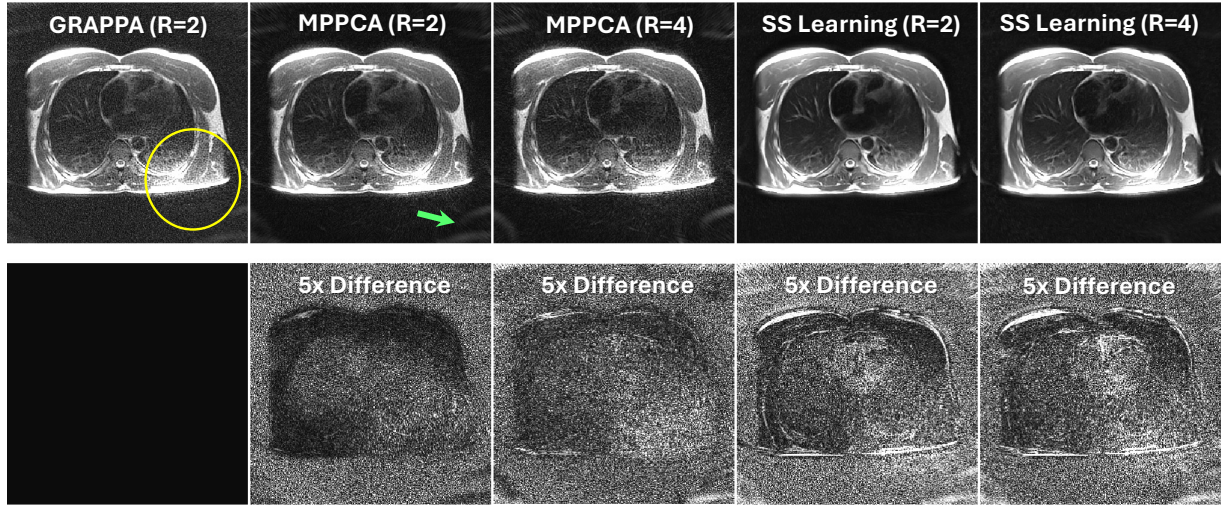

**Figure 3**. Comparison of images reconstructed using conventional GRAPPA, GRAPPA with MPPCA denoising, and the proposed self-supervised learning method with R=2 data and R=4 data, achieved by reducing the number of blades. The difference maps, scaled up by 5 times, shows the absolute difference of the reconstructed image versus the baseline conventional GRAPPA reconstruction without any additional processing. The yellow circle highlights a region of elevated noise, likely due to a failed coil. The green arrow indicates the residual aliasing artifact from the arms due to undersampling. GRAPPA: Generalized autocalibrating partially parallel acquisitions. MPPCA: Marchenko-Pastur principal component analysis.

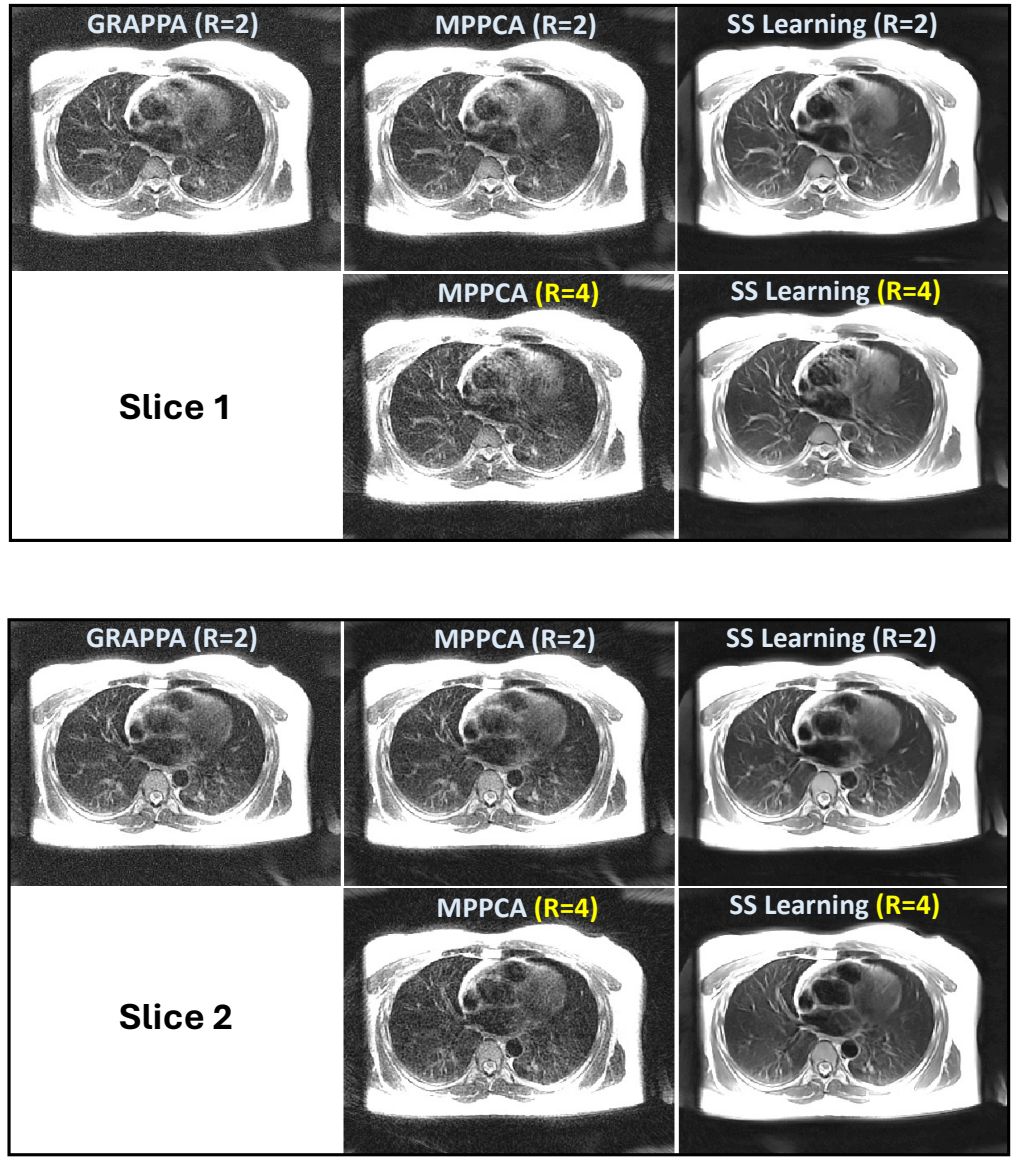

**Figure 4**. Representative axial lung images from two subjects reconstructed using R = 2 data with conventional GRAPPA, GRAPPA with MPPCA denoising, and the proposed self-supervised learning method. Additional reconstructions from retrospectively undersampled R = 4 data, achieved by reducing the number of blades, are shown using GRAPPA with MPPCA denoising and the self-supervised learning method. The self-supervised learning-based reconstructions demonstrate visibly reduced noise and improved suppression of background aliasing artifacts compared to conventional methods. GRAPPA: Generalized autocalibrating partially parallel acquisitions. MPPCA:-–Pastur principal component analysis.

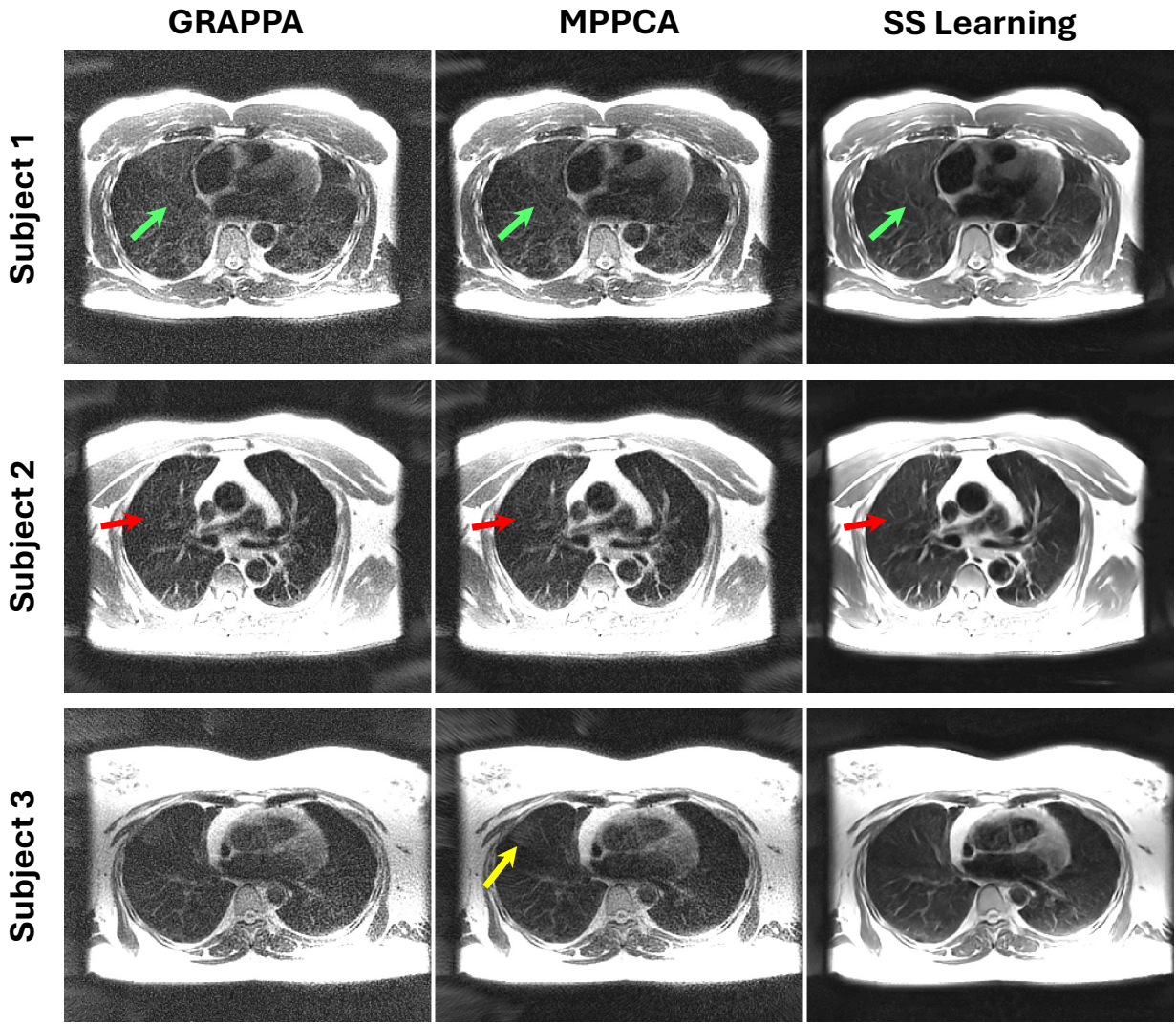

**Figure 5**. Axial lung images comparing GRAPPA reconstruction, GRAPPA with MPPCA denoising, and the proposed self-supervised learning joint denoising and reconstruction for three different subjects, using data acquired at R = 2 acceleration. While MPPCA denoising reduces noise to some extent when applied to GRAPPA reconstructions, its performance is inconsistent. In Subject 1, fine vascular structures are partially recovered after MPPCA denoising, whereas in Subject 2, anatomical details remain unclear. In Subject 3, aliasing artifacts overlap with lung structures in both GRAPPA and GRAPPA+MPPCA reconstructions, hindering interpretation of the parenchyma. In contrast, the self-supervised learning method effectively recovers anatomical detail and suppresses residual aliasing artifacts across all cases. GRAPPA: Generalized autocalibrating partially parallel acquisitions. MPPCA: Marchenko-Pastur principal component analysis.

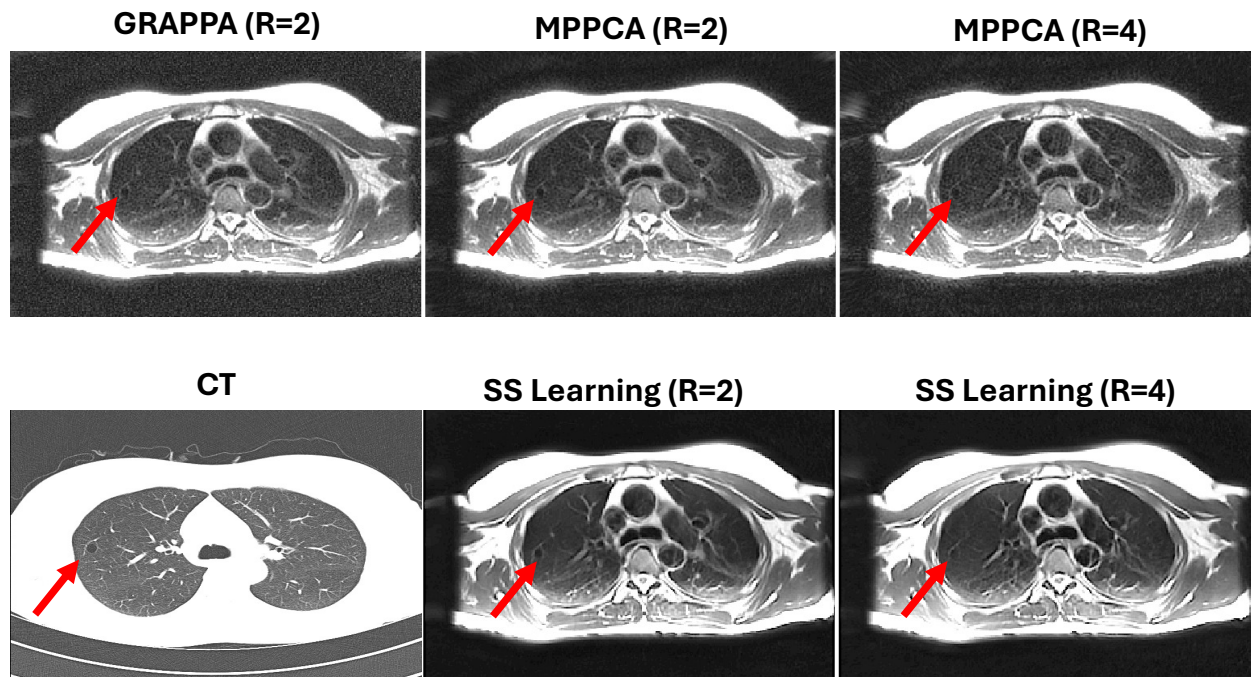

**Figure 6**. Representative axial lung images comparing conventional GRAPPA reconstruction, GRAPPA with MPPCA denoising, and the proposed self-supervised learning joint denoising and reconstruction at acceleration factors of R = 2 and R = 4. The corresponding CT image shows a pneumatocele (thin wall air-filled cyst) in the right upper lobe (red arrow). The proposed self-supervised learning-based method enhances image clarity and improves visualization of the abnormality. GRAPPA: Generalized autocalibrating partially parallel acquisitions. MPPCA: Marchenko-Pastur principal component analysis.

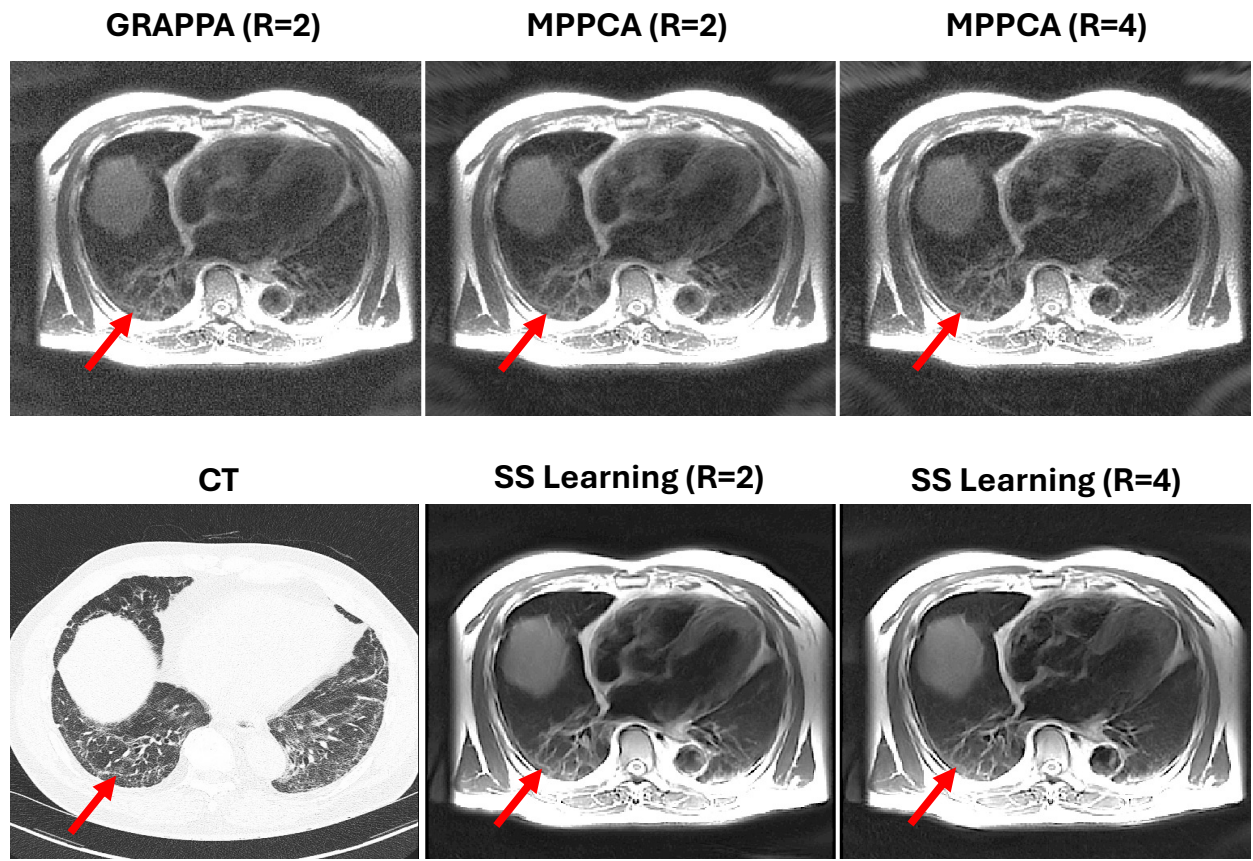

**Figure 7**. Representative axial lung images comparing conventional GRAPPA reconstruction, GRAPPA with MPPCA denoising, and the proposed self-supervised learning joint denoising and reconstruction at acceleration factors of R = 2 and R = 4. The corresponding CT image reveals peripheral ground-glass opacities and subpleural lines (red arrows). These abnormalities are well visualized on the T2-weighted PROPELLER MRI, with the denoised MR images showing good alignment with the CT findings. GRAPPA: Generalized autocalibrating partially parallel acquisitions. MPPCA: Marchenko-Pastur principal component analysis. PROPELLER: Periodically rotated overlapping parallel lines with enhanced reconstruction.

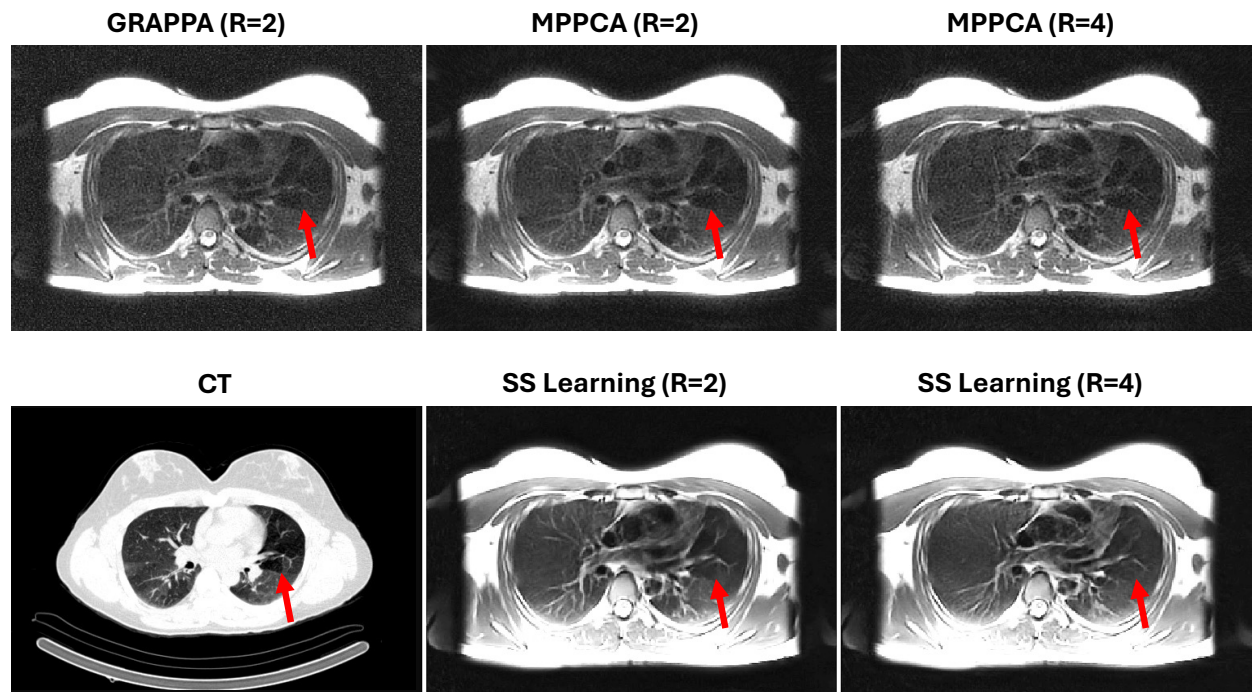

**Figure 8**. Representative axial lung images comparing conventional GRAPPA reconstruction, GRAPPA with MPPCA denoising, and the proposed self-supervised learning joint denoising and reconstruction at acceleration factors of R = 2 and R = 4. The corresponding CT image confirms the presence of multiple perfusion defects (red arrows), which are also visible on the T2-weighted PROPELLER MRI. The self-supervised learning reconstructed images depict the perfusion abnormalities more clearly. GRAPPA: Generalized autocalibrating partially parallel acquisitions. MPPCA: Marchenko-Pastur principal component analysis. PROPELLER: Periodically rotated overlapping parallel lines with enhanced reconstruction.

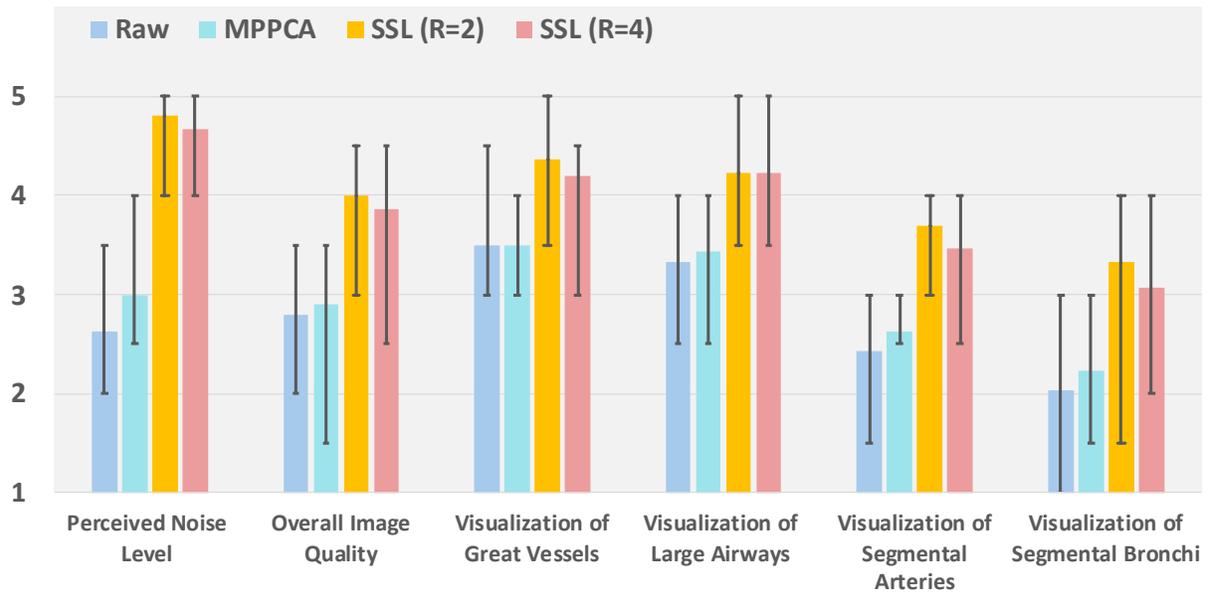

**Figure 9**. Reader scores for image quality assessment across six categories using a five-point Likert scale. The height of each bar represents the mean score across all cases, averaged from two readers; error bars indicate the minimum and maximum scores. Significant improvements were observed for the self-supervised learning-based reconstruction compared to both conventional GRAPPA and GRAPPA with MPPCA denoising. No significant difference was found between self-supervised learning-based reconstructions at R = 2 and R = 4. Detailed statistical results are provided in Supporting Materials Tables S1 and S2. GRAPPA: Generalized autocalibrating partially parallel acquisitions. MPPCA: Marchenko-Pastur principal component analysis.

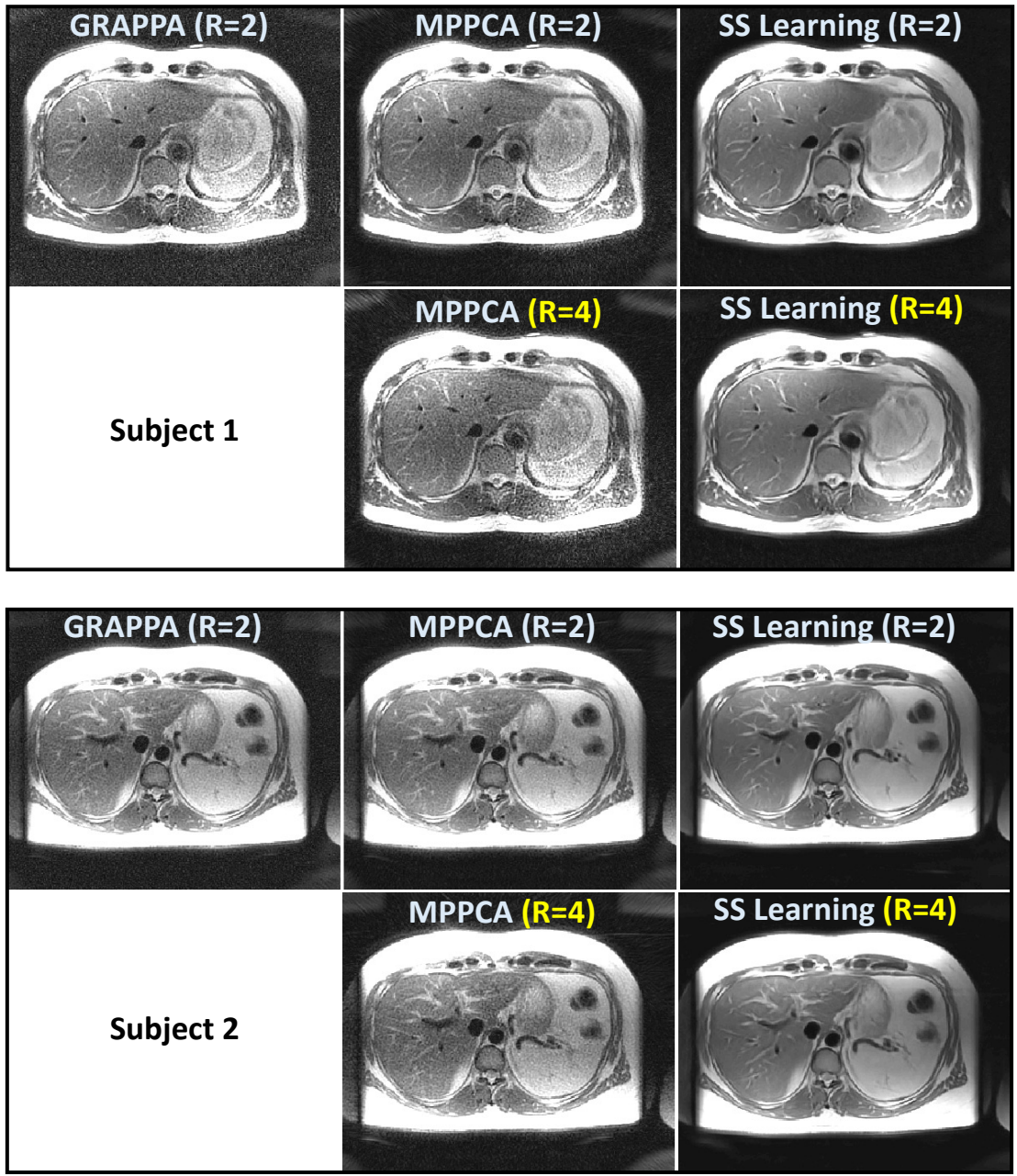

**Figure 10**. Representative axial slices including the liver from two subjects. Images were reconstructed from R = 2 data using conventional GRAPPA, GRAPPA with MPPCA denoising, and the proposed self-supervised learning method. Additional reconstructions from retrospectively undersampled R = 4 data, achieved by reducing the number of blades, are shown using GRAPPA with MPPCA denoising and the self-supervised learning method. The noise level is visually reduced when using the self-supervised learning method and the aliasing in the background is reduced. GRAPPA: Generalized autocalibrating partially parallel acquisitions. MPPCA: Marchenko-Pastur principal component analysis.

**Supporting Materials:**

| Image Quality Scores Two Readers | | Perceived Noise Level | Overall Image Quality | Visualization of Great Vessels | Visualization of Large Airways | Visualization of Segmental Arteries | Visualization of Segmental Bronchovesicular Structures |
|---|---|---|---|---|---|---|---|
| **Mean (SD)** | GRAPPA | 2.63 (0.48) | 2.80 (0.49) | 3.50 (0.46) | 3.33 (0.52) | 2.43 (0.50) | 2.03 (0.58) |
| | MPPCA | 3.00 (0.46) | 2.90 (0.60) | 3.50 (0.33) | 3.43 (0.42) | 2.63 (0.23) | 2.23 (0.46) |
| | SSL (R=2) | 4.80 (0.32) | 4.00 (0.53) | 4.37 (0.35) | 4.23 (0.42) | 3.70 (0.37) | 3.33 (0.75) |
| | SSL (R=4) | 4.67 (0.41) | 3.87 (0.58) | 4.20 (0.46) | 4.23 (0.50) | 3.47 (0.44) | 3.07 (0.56) |
| **Min - Max** | GRAPPA | 2.0 – 3.5 | 2.0 – 3.5 | 3.0 – 4.5 | 2.5 – 4.0 | 1.5 – 3.0 | 1.0 – 3.0 |
| | MPPCA | 2.5 – 4.0 | 1.5 – 3.5 | 3.0 – 4.0 | 2.5 – 4.0 | 2.5 – 3.0 | 1.5 – 3.0 |
| | SSL (R=2) | 4.0 – 5.0 | 3.0 – 4.5 | 3.5 – 5.0 | 3.5 – 5.0 | 3.0 – 4.0 | 1.5 – 4.0 |
| | SSL (R=4) | 4.0 – 5.0 | 2.5 – 4.5 | 3.0 – 4.5 | 3.5 – 5.0 | 2.5 – 4.0 | 2.0 – 4.0 |

**Table S1.** Summary of image quality scores for T2-weighted lung MRI reconstructed using conventional GRAPPA (with and without MPPCA denoising) and the proposed self-supervised learning (SSL)–based joint denoising and reconstruction method, evaluated across six image quality categories. Scores were averaged across two readers. The table reports the mean, standard deviation, minimum, and maximum scores for each method across 15 individual subjects.

| Freidman Test P-value | Perceived Noise Level | Overall Image Quality | Visualization of Great Vessels | Visualization of Large Airways | Visualization of Segmental Arteries | Visualization of Segmental Bronchovesicular Structures |
|---|---|---|---|---|---|---|
| | 9.87e-9 | 3.15e-8 | 6.02e-6 | 6.48e-7 | 1.03e-8 | 1.05e-7 |

**Table S2.** Summary of p-values for the overall test using Freidman test among all four reconstruction methods for each category.

| Wilcoxon signed-rank test  Benjamini-Yekutieli adjusted P-value | Perceived Noise Level | Overall Image Quality | Visualization of Great Vessels | Visualization of Large Airways | Visualization of Segmental Arteries | Visualization of Segmental Bronchovesicular Structures |
|---|---|---|---|---|---|---|
| SSL(R=2) vs GRAPPA | 0.00028 | 0.00027 | 0.00081 | 0.00047 | 0.00031 | 0.00030 |
| SSL(R=2) vs MPPCA | 0.00030 | 0.00046 | 0.00041 | 0.00059 | 0.00030 | 0.00061 |
| SSL(R=4) vs GRAPPA | 0.00030 | 0.00023 | 0.0016 | 0.00067 | 0.00032 | 0.00043 |
| SSL(R=4) vs MPPCA | 0.00030 | 0.00044 | 0.0014 | 0.00089 | 0.00043 | 0.00071 |

**Table S3**. Summary of Benjamini-Yekutieli adjusted p-value for image quality score comparisons using one-tail Wilcoxon sign-rank test of T2-weighted lung MRI between conventional GRAPPA reconstruction (with and without MPPCA denoising) and the proposed self-supervised learning-based joint denoising and reconstruction method.